\documentclass[a4paper,12pt]{article}
\usepackage{amsmath, amssymb, latexsym, amscd, amsthm,amsfonts,amstext}
\usepackage[mathscr]{eucal}
\usepackage{graphicx}
\usepackage{subfig}

\numberwithin{equation}{section}
\input{tcilatex}
\begin{document}

\date{\vspace{-5ex}}
\title{Two reconstruction procedures for a 3-d phaseless inverse scattering
problem for the generalized Helmholtz equation }
\author{Michael V. Klibanov$^{\ast }$ and Vladimir G. Romanov$^{\circ }$
\and $^{\ast }$Department of Mathematics and Statistics \and University of
North Carolina at Charlotte \and Charlotte, NC 28223, USA \and $^{\circ }$%
Sobolev Institute of Mathematics, Novosibirsk 630090, Russian Federation
\and E-mails: mklibanv@uncc.edu and romanov@math.nsc.ru}
\date{}
\maketitle

\begin{abstract}
The 3-d inverse scattering problem of the reconstruction of the unknown
dielectric permittivity in the generalized Helmholtz equation is considered.
The main difference with the conventional inverse scattering problems is
that only the modulus of the scattering wave field is measured.\ The phase
is not measured. The initializing wave field is the incident plane wave. On
the other hand, in the previous recent works of the authors about the
\textquotedblleft phaseless topic" the case of the point source was
considered \cite{KR,KRB,KR2}. Two reconstruction procedures are developed
for a linearized case. However, the linearization is not the Born
approximation. This means that, unlike the Born approximation, our
linearization does not break down when the frequency tends to the infinity.
Applications are in imaging of nanostructures and biological cells.
\end{abstract}

% -------------------------------------------------------

\textbf{AMS classification code:} 35R30.

\graphicspath{{Figures/}}

%
%
% \documentclass{amsart}
% %%%%%%%%%%%%%%%%%%%%%%%%%%%%%%%%%%%%%%%%%%%%%%%%%%%%%%%%%%%%%%
% \usepackage{amssymb}
% \usepackage[T1]{fontenc}
% \usepackage[latin1]{inputenc}
% \usepackage{graphicx}
% \usepackage{geometry}
% \usepackage{url}
% \usepackage{epsfig}
% % \usepackage{labelfig}
% \usepackage{verbatim}
% % \usepackage{umlaut}
% \usepackage{euscript}
% \usepackage{afterpage}
% \usepackage{graphics}
% \usepackage{amsmath}
% \usepackage{pst-plot}
% \usepackage{subfig}
%
% % \input{psfig.sty}
% \newtheorem{theorem}{Theorem}[section]
% \newtheorem{lemma}[theorem]{Lemma}
% \newtheorem{proposition}[theorem]{Proposition}
% \newtheorem{corollary}[theorem]{Corollary}
% \newtheorem{definition}[theorem]{Definition}
% \numberwithin{equation}{section}
% \newcommand{\norm}[1]{\left\Vert#1\right\Vert}
% \newcommand{\abs}[1]{\left\vert#1\right\vert}
% \newcommand{\set}[1]{\left\{#1\right\}}
% \newcommand{\Real}{\mathbb R}
% \newcommand{\eps}{\varepsilon}
% \newcommand{\To}{\longrightarrow}
% \newcommand{\bn}{\mathbf{n}}
% \def\OFEM{\Omega_{FEM}}
% \def\OFDM{\Omega_{FDM}}
% \newcommand\scal[1]{(\!(#1)\!)}
% \newcommand\bscal[1]{\big(\!\big(#1\big)\!\big)}
% % \input{tcilatex}
% \def\bR{\mathbb{R}}
% \def\bx{\mathbf{x}}
%
%
%
% \begin{document}
%
%
%
%
%

%
% \thispagestyle{plain}
%

\graphicspath{{FIGURES/}
{Figures/}
{FiguresJ/newfigures/}
{pics/}}

\section{Introduction}

\label{sec:1}

This paper is the continuation of three recent publications \cite{KR,KRB,KR2}
of the authors dedicated to the reconstruction procedures for 3-d Phaseless
Inverse Scattering Problems (PISPs). Our goal here is to extend the result
of \cite{KR2} to the case when the wave propagation process is generated by
an incident plane wave rather than by the point source of \cite{KR2}. We end
up our derivation with the inversion of the 2-d Radon transform.\ This can
be done by the well known technique, see, e.g. \cite{Nat}. We also propose a
different inversion method based on the integral geometry. The idea of the
second method goes back to the famous paper of Allan Cormack published in
1963 \cite{Cormack}. It is well known that his publication \cite{Cormack}
led him to the Nobel prize in 1979, along with Godfrey Hounsfield, for his
work on X-ray computed tomography (CT), see
https://en.wikipedia.org/wiki/Allan\_McLeod\_Cormack.

In \cite{KR} a reconstruction procedure was proposed for a 3-d PISP for the
Schr\"{o}dinger equation in the frequency domain.\ The next question is
about the reconstruction for the generalized Helmholtz equation. An
important difference between these two is that in the generalized Helmholtz
equation the unknown coefficient is multiplied $k^{2},$ where $k>0$ is the
frequency, and $k$ is varied in our problems. On the other hand, in the Schr%
\"{o}dinger equation the unknown potential is not multiplied by $k^{2}.$
This makes the reconstruction procedure for the PISP for the Schr\"{o}dinger
equation easier than that for the generalized Helmholtz equation.

In \cite{KRB} the reconstruction procedure was developed for the case when
the Born approximation is considered for the generalized Helmholtz equation.
However linearization of the Born approximation breaks down for large values
of $k$. Hence, in \cite{KR2} the reconstruction procedure for that equation
was developed for the case when the wave propagation process is generated by
the point source. Even though a linearization was used in \cite{KR2}, this
linearization does not break down for large $k$. Still, we point out that
both here and in \cite{KR2} the phase of the scattered wave field is
reconstructed without the linearization, separately for each direction of
the incident wave field.\ The only approximation we use when reconstructing
the phase is that we ignore the term $O\left( 1/k\right) $ for $k\rightarrow
\infty .$

Inverse scattering problems without the phase information arise in imaging
of structures whose sizes are of the micron range or less.\ Recall that 1
micron ($1\mu m)=10^{-6}m.$ For example, nano structures typically have
sizes of hundreds on nanometers, which is about $0.1\mu m.$\ Therefore, the
wavelength for this imaging should be about $0.1\mu m.$ The second example
is in imaging of living biological cells. Sizes of cells are between $5\mu m$
and $100\mu m$ \cite{PM,Bio}. Either optical of X-ray radiation is used in
this imaging. It is well known that for the micron range of wavelengths only
the intensity of the scattered wave field can be measured, and the phase
cannot be measured \cite{Dar,Die,Khach,Pet,Ruhl}. The intensity is the
square modulus of the scattered complex valued wave field. We point out that
standard statements of inverse scattering problems assume that both the
modulus and the phase of the complex valued scattered wave field can be
measured, see, e.g. \cite{Hu,Is,Li,Nov1,Nov2}.

As to the uniqueness results PISPs, see \cite{AS,KS,NHS} for the 1-d case
and \cite{KSIAP,AML,AA} for the 3-d case. However, proofs of these results
are not constructive. The first rigorous reconstruction procedures were
proposed recently by the authors \cite{KR,KRB,KR2} and independently and for
different statements of problems by Novikov \cite{Nov3,Nov4}. In \cite%
{Iv1,Iv2} numerical methods for reconstructions of shapes of obstacles from
phaseless scattered data were developed. As to some heuristic reconstruction
algorithms, we refer to, e.g. works of physicists \cite%
{Dar,Die,Khach,Pet,Ruhl}.

In section 2 we state both forward and inverse problems. In section 3 we
study geodesics lines generated by the refractive index $n(x),x\in \mathbb{R}%
^{3}.$ In section 4 we study an auxiliary Cauchy problem for a hyperbolic
equation. In section 5 we show how to reconstruct the phase for high
frequencies. In section 6 we linearize the problem. Finally in section 7 we
present two reconstruction procedures.

\section{Statements of forward and inverse problems}

\label{sec:2}

Let $B>0$ and $\Omega =\{|x|<B\}\subset \mathbb{R}^{3}$ be the ball of the
radius $B$ with the center at $\{x=0\}.$ Denote $S=\left\{ \left\vert
x\right\vert =B\right\} .$ Let the refractive index $n(x),x\in \mathbb{R}%
^{3} $ be a function satisfying the following conditions 
\begin{equation}
n(x)\in C^{15}(\mathbb{R}^{3}),\quad n(x)=1+\beta (x),  \label{1.0}
\end{equation}%
\begin{equation}
\beta (x)\geq 0,\text{ }\beta (x)=0\quad \text{for }\>x\in \mathbb{R}%
^{3}\setminus \Omega .  \label{1.3}
\end{equation}%
The $C^{15}$ smoothness of $n(x)$ will be explained below (see the proof of
Theorem 1). Let $S^{2}=\{\nu \in \mathbb{R}^{3}:\>|\nu |=1\}$ be the unit
sphere. We consider the following equation 
\begin{equation}
\Delta u+k^{2}n^{2}(x)u=0,\quad x\in \mathbb{R}^{3},  \label{1.4}
\end{equation}%
$u\left( x,k\right) $ is the complex valued wave field and $k>0$ is a
positive real number. We seek the solution of equation (\ref{1.4}) in the
form%
\begin{equation}
u\left( x,k,\nu \right) =\exp \left( -ikx\cdot \nu \right) +u_{sc}\left(
x,k,\nu \right) ,  \label{1.4a}
\end{equation}%
where the function $u_{sc}\left( x,k,\nu \right) $ satisfies the radiation
condition at the infinity, 
\begin{equation}
\frac{\partial u_{sc}}{\partial r}+iku_{sc}=o(r^{-1}),\>\>r=|x|\rightarrow
\infty .  \label{1.4b}
\end{equation}%
It is well known that the problem (\ref{1.4})-(\ref{1.4b}) has unique
solution $u\left( x,k,\nu \right) \in C^{3}\left( \mathbb{R}^{3}\right)
,\forall \nu \in S^{2}$ \cite{CK}.\ Furthermore, theorem 6.17 of \cite{GT}
implies that, given (\ref{1.0}), the function $u\left( x,k,\nu \right) \in
C^{16+\alpha }\left( \mathbb{R}^{3}\right) ,\forall \alpha \in \left(
0,1\right) ,\forall \nu \in S^{2},$ where $C^{16+\alpha }\left( \mathbb{R}%
^{3}\right) $ is the H\"{o}lder space.

We model the propagation of the electric wave field in $\mathbb{R}^{3}$ by
the solution of the problem (\ref{1.4})- (\ref{1.4b}). This modeling was
numerically justified in \cite{BMM} in the case of the time domain.\ It was
shown numerically in \cite{BMM} that this modeling can replace the modeling
via the full Maxwell's system, provided that only a single component of the
electric field is incident upon the medium. Then this component dominates
two others and its propagation is well governed by the single equation (\ref%
{1.4}). This conclusion was verified via accurate imaging using
electromagnetic experimental data in, e.g. Chapter 5 of \cite{BK} and \cite%
{TBKF1,TBKF2}.

We consider the following inverse problem:

\textbf{Phaseless Inverse Scattering Problem.} \emph{Consider a number }$%
k_{0}>0$\emph{. Find the function }$\beta \left( x\right) $\emph{\ assuming
that the following function }$f\left( x,\nu ,k\right) $\emph{\ is given}%
\begin{equation}
f\left( x,\nu ,k\right) =\left\vert u_{sc}\left( x,\nu ,k\right) \right\vert
^{2},\>\forall x\in S,\forall \nu \in S^{2},\>\forall k\geq k_{0},
\label{100}
\end{equation}

\section{Geodesic lines}

\label{sec:3}

Our reconstruction procedure is essentially using geodesic lines generated
by the function $n(x).$Hence, we consider these lines in this section. The
Riemannian metric generated by the function $n(x)$ is 
\begin{equation}
d\tau =n(x)\left\vert dx\right\vert ,|dx|=\sqrt{%
(dx_{1})^{2}+(dx_{2})^{2}+(dx_{3})^{2}}.  \label{3.1}
\end{equation}%
For each vector $\nu \in S^{2}$ define the plane $\Sigma (\nu )$ as%
\begin{equation}
\Sigma (\nu )=\{\xi \in \mathbb{R}^{3}:\>\xi \cdot \nu =-B\}.  \label{3.100}
\end{equation}%
Let $\xi _{0}\in \Sigma (\nu )$ be an arbitrary point. Then $\left( \xi -\xi
_{0}\right) \cdot \nu =0,\forall \xi \in \Sigma (\nu ).$ Hence, the plane $%
\Sigma (\nu )$ is orthogonal to the vector $\nu .$ Also, note that 
\begin{equation}
\Sigma (\nu )\cap \Omega =\varnothing .  \label{3.11}
\end{equation}%
We can represent an arbitrary point $\xi _{0}\in \Sigma (\nu )$ as 
\begin{equation*}
\xi _{0}=\xi _{0}(\eta _{1},\eta _{2})=-B\nu +e_{1}\eta _{1}+e_{2}\eta
_{2},\quad (\eta _{1},\eta _{2})\in \mathbb{R}^{2},
\end{equation*}%
where unit vectors $\nu $, $e_{1}$, $e_{2}$ form an orthogonal triple.

Let the function $\tau (\xi ,\nu )$ be the solution of the Cauchy problem
for the eikonal equation, 
\begin{equation}
|\nabla _{\xi }\tau (\xi ,\nu )|=n(\xi ),\quad \tau |_{\Sigma (\nu )}=0,
\label{3.2}
\end{equation}%
such that 
\begin{equation}
\tau (\xi ,\nu )\left\{ 
\begin{array}{c}
>0\text{ if }\xi \cdot \nu >-B, \\ 
<0\text{ if }\xi \cdot \nu <-B.%
\end{array}%
\right.  \label{3.3}
\end{equation}

It is well known that $|\tau (\xi ,\nu )|$ is the Riemannian distance
between the point $\xi $ and the plane $\Sigma (\nu )$. Physically, $|\tau
(\xi ,\nu )|$ is the travel time between $\xi $ and the plane $\Sigma (\nu )$%
. For $\xi \cdot \nu <-B$ the function $\tau (\xi ,\nu )$ has the form $\tau
(\xi ,\nu )=\xi \cdot \nu +B$. To find this function for $\xi \cdot \nu >-B,$
we need to solve the problem (\ref{3.2}) in this domain. It is well known
that to solve this problem, one needs to solve a system of ordinary
differential equations. These equations also define geodesic lines of the
Riemannian metric. They are (see, e.g. \cite{R3}): 
\begin{equation}
\frac{d\xi }{ds}=\frac{p}{n^{2}(\xi )},\quad \frac{dp}{ds}=\nabla \left( \ln
n(\xi )\right) ,\quad \frac{d\tau }{ds}=1,  \label{1.3a}
\end{equation}%
where $s$ is a parameter and $p=\nabla _{\xi }\tau (\xi ,\nu )$. Consider an
arbitrary point $\xi _{0}(\eta _{1},\eta _{2})\in \Sigma (\nu )$ and the
solution of the equations (\ref{1.3a}) with the Cauchy data 
\begin{equation}
\xi |_{s=0}=\xi _{0}(\eta _{1},\eta _{2}),\quad p|_{s=0}=n(\xi _{0}(\eta
_{1},\eta _{2}))\nu ,\quad \tau |_{s=0}=0.  \label{1.3b}
\end{equation}%
Here the parameter $s\geq 0.$ The solution of the problem (\ref{1.3a}), (\ref%
{1.3b}) defines the geodesic line which passes trough the point $\xi
_{0}(\eta _{1},\eta _{2})$ in the direction $\nu $. Hence, this line
intersects the plane $\Sigma (\nu )$ orthogonally. For $s>0$ the solution
determines the geodesic line $\xi =g(s,\eta _{1},\eta _{2},\nu )$ and the
vector $p=h(s,\eta _{1},\eta _{2},\nu )$ which lays in the tangent direction
to the geodesic line. It is well known from the theory of Ordinary
Differential Equations that if the function $n(x)\in C^{m}(\mathbb{R}^{3})$, 
$m\geq 2$, then functions $g,h$ are $C^{m-1}-$smooth functions of their
variables.

By (\ref{1.3b}) 
\begin{equation*}
\frac{\partial \xi }{\partial \eta _{1}}\Big|_{s=0}=e_{1},\quad \frac{%
\partial \xi }{\partial \eta _{2}}\Big|_{s=0}=e_{2}.
\end{equation*}%
Moreover, it follows from (\ref{1.3a}) and (\ref{1.3b}) that 
\begin{equation*}
\frac{d\xi }{ds}\Big|_{s=0}=\frac{\nu }{{n(\xi _{0}(\eta _{1},\eta _{2}))}}%
=\nu .
\end{equation*}
Hence, 
\begin{equation}
\left\vert \frac{\partial (\xi _{1},\xi _{2},\xi _{3})}{\partial (s,\eta
_{1},\eta _{2})}\right\vert _{s=0}=1\neq 0.  \label{1.3c}
\end{equation}

By (\ref{1.3c}) the equality $x=g(s,\eta _{1},\eta _{2},\nu )$ can be
uniquely solved with respect to $s,\eta _{1},\eta _{2}$ for those points $x$
which are sufficiently close to the plane $\Sigma (\nu )$, as 
\begin{equation*}
s=s(x,\nu ),\quad \eta _{1}=\eta _{1}(x,\nu ),\quad \eta _{2}=\eta
_{2}(x,\nu ).
\end{equation*}%
Hence, the equation 
\begin{equation*}
\xi =g(s,\eta _{1}(x,\nu ),\eta _{2}(x,\nu ),\nu )=\hat{g}(s,x,\nu ),\>s\in %
\left[ 0,s(x,\nu )\right]
\end{equation*}%
defines the geodesic line $\Gamma (x,\nu )$ that passes through points $x$
and $\xi _{0}(\eta _{1}(x,\nu ),\eta _{2}(x,\nu ))=\xi _{0}(x,\nu )$.\
Extend $\Gamma (x,\nu )$ for $\xi \cdot \nu <-B$ by the equation $\xi =\xi
_{0}(x,\nu )+s\nu $, $s<0$. This line intersects the plane $\Sigma (\nu )$
at the point $\xi _{0}(x,\nu )\in \Sigma (\nu )$ orthogonally. The
Riemannian distance between points $x$ and $\xi _{0}(x,\nu )$ is $s(x,\nu
)=\tau (x,\nu )$. Note that functions $\hat{g}(s,x,\nu )$ and $\hat{h}%
(s,x,\nu )=h(s,\eta _{1}(x,\nu ),\eta _{2}(x,\nu ),\nu )$ are $C^{m-1}-$%
smooth functions of their arguments. Since, moreover, $\hat{h}(s(x,\nu
),x,\nu )=\nabla _{x}\tau (x,\nu )$, we conclude that $\tau (x,\nu )$ is $%
C^{m}$-smooth function. In our case $\tau (x,\nu )$ is $C^{15}$-smooth
function of $x$ for $x$ enough closed to $\Sigma (\nu )$.

We have constructed geodesic lines $\Gamma (x,\nu )$\ above only
\textquotedblleft locally", i.e. only for those points $x$\ which are
located sufficiently close to the plane $\Sigma (\nu ).$\ However, we need
to consider these lines \textquotedblleft globally". Hence, everywhere below
we rely on the following Assumption:

\textbf{Assumption}. \emph{We assume that geodesic lines of the metric (\ref%
{3.1}) with conditions (\ref{3.2}), (\ref{3.3}) satisfy the regularity
condition in }$\mathbb{R}^{3}$\emph{. In other words, for each vector }$\nu
\in S^{2}$ \emph{\ and for each point }$x\in \mathbb{R}^{3}$ \emph{there
exists a single geodesic line }$\Gamma \left( x,\nu \right) $ \emph{%
connecting} $x$\emph{\ with the plane } $\Sigma (\nu )$ \emph{such that } $%
\Gamma \left( x,\nu \right) $ \emph{\ intersects }$\Sigma (\nu )$\emph{\
orthogonally. }

It is well known from the Hadamard-Cartan theorem \cite{Ball} that in any
simply connected complete manifold with a non positive curvature each two
points can be connected by a single geodesic line. The manifold $(\Omega ,n)$
is called the manifold of a non positive curvature, if the section
curvatures $K(x,\sigma )\leq 0$ for all $x\in \Omega $ and for all
two-dimensional planes $\sigma $. A sufficient condition for $K(x,\sigma
)\leq 0$ was derived in \cite{R4} 
\begin{equation}
\sum_{i,j=1}^{3}\frac{\partial ^{2}\ln n(x)}{\partial x_{i}\partial x_{j}}%
\xi _{i}\xi _{j}\geq 0,\>\forall x\in \Omega ,\>\xi \in \mathbb{R}^{3}.
\label{1.3d}
\end{equation}%
In our case if condition (\ref{1.3d})\emph{\ }holds for $x\in \Omega $, then
is also holds for all $x\in \mathbb{R}^{3}.$

Our Assumption implies that for any $\nu \in S^{2}$ and for any $x\in 
\mathbb{R}^{3}$ there exists a unique pair $\left( \xi _{0}(x,\nu ),s(x,\nu
)\right) $ with $\xi _{0}(x,\nu )\in \Sigma (\nu )$ such that the geodesic
line $\Gamma (x,\nu )$ connects points $x$ and $\xi _{0}(x,\nu )$ and
intersects the plane $\Sigma (\nu )$ orthogonally. Hence, there exists a
function $\overline{g}(s,x,\nu )$ such that the equation of the geodesic
line $\Gamma (x,\nu )$ is given by 
\begin{equation*}
\Gamma (x,\nu )=\{\xi :\>\xi =\overline{g}(s,x,\nu ),s\in \lbrack 0,s(x,\nu
)]\}
\end{equation*}%
and $x=\overline{g}(s(x,\nu ),x,\nu )$. Then $s(x,\nu )=\tau (x,\nu )$ is
the Riemannian length of $\Gamma (x,\nu )$. Also, $\hat{p}(s(x,\nu ),x,\nu
)=\nabla _{x}\tau (x,\nu )$ and $|\nabla _{x}\tau (x,\nu )|=n(x)$.

\section{Auxiliary Cauchy problem for a hyperbolic equation}

\label{sec:4}

Modifying ideas of \cite{KR,KRB,KR2}, we consider in this section an
auxiliary Cauchy problem for a hyperbolic equation with the incident plane
wave. Consider the hyperbolic equation 
\begin{equation}
n^{2}(x)v_{tt}-\Delta v=0,\quad x\in \mathbb{R}^{3},t\in \mathbb{R}.
\label{5.1}
\end{equation}%
We seek the solution $v\left( x,t,\nu \right) $ of equation (\ref{5.1}) in
the form\textbf{\ }%
\begin{equation}
v\left( x,t,\nu \right) =\delta (t-x\cdot \nu )+\overline{v}\left( x,t,\nu
\right) ,  \label{5.2}
\end{equation}%
where $\delta (t-x\cdot \nu )$ is the incident plane wave propagating in the
direction $\nu ,$ and the function $\overline{v}\left( x,t,\nu \right) $
satisfies the following initial conditions%
\begin{equation}
\overline{v}\mid _{t<-B}\equiv 0.  \label{5.3}
\end{equation}%
Let $T>0$ be an arbitrary number. Let $\varphi (x,\nu )$ be the solution of
the following problem 
\begin{equation}
|\nabla _{x}\varphi (x,\nu )^{2}|=n^{2}(x),\quad  \label{5.4}
\end{equation}%
\begin{equation}
\varphi (x,\nu )=x\cdot \nu \,\text{\ for all }\>x\cdot \nu \leq -B.
\label{5.40}
\end{equation}%
Let the function $A(x,\nu )$ be 
\begin{equation}
A(x,\nu )=\left\{ 
\begin{array}{rl}
\exp \left( -\frac{1}{2}\int\limits_{\Gamma (x,\nu )}n^{-2}(\xi )\Delta
_{\xi }\varphi (\xi ,\nu )d\tau \right) , & x\cdot \nu >-B, \\ 
1, & x\cdot \nu \leq -B.%
\end{array}%
\right.  \label{5.5}
\end{equation}%
The reason of the second line of (\ref{5.5}) is (\ref{5.40}) as well as the
fact that $\Gamma (x,\nu )$ is the straight line for $x\cdot \nu \leq -B.$
Let $T>0$ be an arbitrary number. Consider the domain $D(\nu ,T)$ defined as%
\begin{equation}
D(\nu ,T)=\{(x,t)|\>\max (-B,\varphi (x,\nu ))\leq t\leq T\}.  \label{5.6}
\end{equation}%
Let $H\left( t\right) $ be the Heaviside function,%
\begin{equation*}
H\left( t\right) =\left\{ 
\begin{array}{c}
1,t>0, \\ 
0,t<0.%
\end{array}%
\right.
\end{equation*}

\textbf{Theorem 1. }\emph{Let conditions (\ref{1.0}), (\ref{1.3}) as well as
Assumption hold. Let }$\nu \in S^{2}$\emph{\ be an arbitrary vector and }$T$ 
\emph{\ be an arbitrary number. Then there exists unique solution }$%
v(x,t,\nu )$\emph{\ of the problem (\ref{5.1})-(\ref{5.3}) which can be
represented in the form }%
\begin{equation}
v(x,t,\nu )=A(x,\nu )\delta (t-\varphi (x,\nu ))+\widehat{v}(x,t,\nu
)H(t-\varphi (x,\nu )),x\in \mathbb{R}^{3},t\in \left( -\infty ,T\right) ,
\label{5.7}
\end{equation}%
\emph{where the function }$A(x,\nu )$\emph{\ is defined in (\ref{5.5}), the
function }$\widehat{v}(x,t,\nu )\in C^{2}(D(\nu ,T))$\emph{\ and }$\widehat{v%
}(x,t,\nu )=0$\emph{\ for} $t<-B$.

\textbf{Proof}. Introduce functions $\theta _{k}(t)$ as%
\begin{equation*}
\theta _{-3}(t)=\delta ^{\prime \prime }(t),\text{ }\theta _{-2}(t)=\delta
^{\prime }(t),\text{ }\theta _{-1}(t)=\delta (t),\text{ }\theta _{k}(t)=%
\frac{t^{k}}{k!}H(t),\quad k=0,1,2,\ldots
\end{equation*}%
Observe that $\theta _{k}^{\prime }(t)=\theta _{k-1}(t)$ for all $k\geq -2$.
We seek the solution to the problem (\ref{5.1})-(\ref{5.3})\emph{\ }in the
form 
\begin{equation}
v(x,t,\nu )=A(x,\nu )\theta _{-1}\left( t-\varphi (x,\nu )\right)
+\sum\limits_{k=0}^{r}a_{k}(x,\nu )\theta _{k}(t-\varphi (x,\nu
))+w_{r}(x,t,\nu )  \label{4.1}
\end{equation}%
for $x\in \mathbb{R}^{3},t\in \mathbb{R},r\geq 1.$ Substituting
representation (\ref{4.1}) in (\ref{5.1}), using $|\nabla _{x}\varphi (x,\nu
)^{2}|=n^{2}(x)$ and equating coefficients at $\theta _{k}(t)$, we obtain
formulas for finding coefficients $A(x,\nu )$, $a_{k}(x,\nu )$: 
\begin{eqnarray}
&&2\nabla A(x,\nu )\cdot \nabla \varphi (x,\nu )+A(x,\nu )\Delta \varphi
(x,\nu )=0,  \notag \\
&&2\nabla a_{0}(x,\nu )\cdot \nabla \varphi (x,\nu )+a_{0}(x,\nu )\Delta
\varphi (x,\nu )=\Delta A(x,\nu ),  \label{4.2} \\
&&2\nabla a_{k}(x,\nu )\cdot \nabla \varphi (x,\nu )+a_{k}(x,\nu )\Delta
\varphi (x,\nu )=\Delta a_{k-1}(x,\nu ),\>k\geq 1.\qquad  \notag
\end{eqnarray}%
Since by (\ref{5.2}) and (\ref{5.3}) $v\left( x,t,\nu \right) =\delta
(t-x\cdot \nu )$ for $t<-B$, then we obtain 
\begin{equation}
A(x,\nu )=1,\>\text{ for }\>x\cdot \nu \leq -B,  \label{4.3}
\end{equation}%
\begin{equation}
\varphi (x,\nu )=x\cdot \nu ,\text{ for }\>x\cdot \nu \leq -B,  \label{4.3a}
\end{equation}%
\begin{equation}
a_{k}(x,\nu )=0,\>k\geq 0,\text{ for }\>x\cdot \nu \leq -B.  \label{4.3b}
\end{equation}

Moreover using (\ref{5.1})-(\ref{5.3}) and (\ref{4.1})-(\ref{4.3}), we
obtain the following Cauchy problem for the residual $w_{s}(x,t,\nu )$ of
the expansion (\ref{4.1}) 
\begin{equation}
n^{2}(x)\partial _{t}^{2}w_{r}-\Delta w_{r}=F_{r}(x,t,\nu ),  \label{4.4}
\end{equation}%
\begin{equation}
w_{r}|_{t<-B}=0,  \label{4.5}
\end{equation}%
\begin{equation}
F_{r}(x,t,\nu )=(\Delta a_{r}(x,\nu ))\theta _{r}(t-\varphi (x,\nu )).
\label{4.6}
\end{equation}

We now construct solutions of equations (\ref{4.2}) with the Cauchy data (%
\ref{4.3})-(\ref{4.3b}). We rewrite equations (\ref{1.3a}) of the geodesic
line $\Gamma (x,\nu )$ with the vector $p(x,\nu )=\nabla \varphi (x,\nu )$
in the following form 
\begin{equation}
\frac{dx}{d\tau }=n^{-2}(x)p(x,\nu ),\quad \frac{dp(x,\nu )}{d\tau }=\nabla
\left( \ln n(x)\right) ,  \label{4.7}
\end{equation}%
where $\tau $ is the arc Riemannian length with the element length given by
the formula $d\tau =n(\xi )|d\xi |$. Hence, we have along the geodesic line $%
\Gamma (x,\nu )$ 
\begin{equation*}
2\nabla A(x,\nu )\cdot \nabla \varphi (x,\nu )=2\nabla A(x,\nu )\cdot
p(x,\nu )=2n^{2}(x)\nabla A(x,\nu )\cdot \frac{dx}{d\tau }=2n^{2}(x)\frac{%
dA(x,\nu )}{d\tau }.
\end{equation*}%
Hence, the first equation (\ref{4.2}) becomes 
\begin{equation}
2n^{2}(x)\frac{dA(x,\nu )}{d\tau }+A(x,\nu )\Delta \varphi (x,\nu )=0.
\label{4.7a}
\end{equation}%
The solution of this equation with the initial condition (\ref{4.3}) is
given by (\ref{5.5}).

Then equations for $a_{k}(x,\nu )$, $k\geq 0$, can be rewritten in the form 
\begin{equation}
\frac{d}{d\tau }\Big(\frac{a_{k}(x,\nu )}{A(x,\nu )}\Big)=\frac{\Delta
a_{k-1}(\xi ,\nu )}{2n^{2}(x)A(x,\nu )},\quad k\geq 0,  \label{4.70}
\end{equation}%
where we should formally have $a_{-1}(x,\nu )=A(x,\nu )$. It is easy to
verify (\ref{4.70}), if expressing $\Delta \varphi (x,\nu )$ via $A(x,\nu )$
using (\ref{4.7a}) .

Taking into account the initial data (\ref{4.3b}), we obtain 
\begin{equation}
a_{k}(x,\nu )=A(x,\nu )\int\limits_{\Gamma (x,\nu )}\frac{\Delta a_{k-1}(\xi
,\nu )}{2n^{2}(\xi )A(\xi ,\nu )}d\tau ,\>k\geq 0,\quad x\cdot \nu \geq -B.
\label{4.9}
\end{equation}

Let $m>1$ be a sufficiently large integer which we will choose below. If $%
n(x)\in C^{m}(\mathbb{R}^{3}),$ then we have 
\begin{equation*}
\varphi (x,\nu )\in C^{m}(\mathbb{R}^{3}),A(x,\nu )\in C^{m-2}(\mathbb{R}%
^{3}),a_{k}(x,\nu )\in C^{m-4-2k}(\mathbb{R}^{3}),a_{r}(x,\nu )\in
C^{m-4-2r}(\mathbb{R}^{3}).
\end{equation*}%
Hence, the function $F_{r}(x,t,\nu )\in C^{l}(D(\nu ,T))$, where $l=\min
(m-6-2r,r).$ By (\ref{4.3b}) and (\ref{4.6}) 
\begin{equation}
F_{r}(x,t,\nu )=0\text{ for }t\leq \varphi (x,\nu )\text{ and for }x\cdot
\nu <-B.  \label{4.91}
\end{equation}

We now prove that%
\begin{equation}
w_{r}\left( x,t,\nu \right) =0\text{ for }t\in \left[ -B,\varphi \left(
x,\nu \right) \right] .  \label{4.10}
\end{equation}%
Indeed, (\ref{4.5}) implies that (\ref{4.10}) is true for $t\leq -B$. Let
now $(x^{0},t^{0})$ be an arbitrary point of the domain $Q(\nu )=\{(x,t)\in
R^{4}:\,-B<t<\varphi (x,\nu )\}$. Let $\tau (x,x^{0})$ be the Riemannian
distance between points $x$ and $x^{0}$. Denote 
\begin{equation*}
K(\nu ,x^{0})=\{(x,t)\in \mathbb{R}^{4}|\,-B\leq t\leq \varphi (x^{0},\nu
)-\tau (x,x^{0})\}\mathbf{.}
\end{equation*}%
Hence, $K(\nu ,x^{0})$ is the inner part of the characteristic cone with the
vertex at the point $(x^{0},\varphi (x^{0},\nu ))$ and bounded by the plane $%
t=-B$. Obviously $(x^{0},t^{0})\in K(\nu ,x^{0})$. By (\ref{4.91}) $%
F_{r}(x,t,\nu )=0$ for $(x,t)\in K(\nu ,x^{0})$. Apply now the method of
energy estimates to the problem (\ref{4.4}), (\ref{4.5}) in the domain $%
K(\nu ,x^{0})$. Observe that the resulting surface integral will be
non-negative on the lateral boundary $t=\varphi (x^{0},\nu )-\tau (x,x^{0})$
of the characteristic cone $K(\nu ,x^{0}).$ Also, the integral over $\left\{
t=-B\right\} \cap \overline{K}(\nu ,x^{0})$ will be zero due to (\ref{4.5}).
Hence, we obtain $w_{r}(x,t,\nu )\equiv 0$ for all $(x,t)\in K(\nu ,x^{0})$.
Thus, $w_{r}(x^{0},t^{0},\nu )=0$. Since $(x^{0},t^{0})$ is an arbitrary
point of the domain $Q(\nu ),$ then we conclude that (\ref{4.10}) holds.

Furthermore, using theorems 3.2, 4.1, corollary 4.2 and energy estimates of
Chapter 4 of \cite{Lad}, one can easily prove that there exists unique
solution $w_{r}$ of equation (\ref{4.4}), (\ref{4.5}) such that 
\begin{equation*}
w_{r}(x,-B,\nu )=\partial _{t}w_{r}(x,-B,\nu )=0
\end{equation*}%
and $w_{r}(x,t,\nu )\in H^{l+1}(Y(t,\nu ,T))$, $\partial _{t}w_{r}(\cdot
,t,\nu )\in H^{l}(Y(t,\nu ,T))$ for all $t\in (-B,T)$, where $Y(t,\nu
,T)=D(\nu ,T)\cap \{t=const\}$. Choosing $m=15$ and $r=3$, we obtain $l=3$.
Thus, $w_{3}(x,t,\nu )\in H^{4}(Y(t,\nu ,T))$ and $\partial
_{t}w_{3}(x,t,\nu )\in H^{3}(Y(t,\nu ,T))$. Therefore the embedding theorem
implies that $w_{3}\in C^{2}\left( D(\nu ,T)\right) $.

Setting 
\begin{equation*}
\widehat{v}(x,t,\nu )=\sum\limits_{k=0}^{3}a_{k}(x,\nu )\frac{(t-\varphi
(x,\nu ))^{k}}{k!}+w_{3}(x,t,\nu ),
\end{equation*}%
we obtain (\ref{5.7}) as well as the required smoothness $\widehat{v}%
(x,t,\nu )\in C^{2}(D(\nu ,T))$. $\square $\textbf{\ }

\subsection{Connection with the problem (\protect\ref{1.4})-(\protect\ref%
{1.4b})}

Let $v(x,t,\nu )$ be the solution of the problem (\ref{5.1})-(\ref{5.3})
which is guaranteed by Theorem 1. Fix an arbitrary bounded domain $\Phi
\subset \mathbb{R}^{3}$. And consider the behavior of the functions $%
\partial _{t}^{k}v(x,t,\nu ),k=0,1,2$ for $x\in \Phi $ and $t\rightarrow
\infty $. We refer to Theorem 4 of Chapter 10 of the book \cite{V} as well
as to Remark 3 after that theorem. It follows from these results that these
functions decay exponentially as $t\rightarrow \infty $ as long as $x\in
\Phi $.

Consider the Fourier transform $V\left( x,k,y\right) $ of the function $v,$%
\begin{equation}
V(x,k,\nu )=\dint\limits_{-\infty }^{\infty }v\left( x,t,\nu \right) \exp
\left( -ikt\right) dt,\quad x\in G.  \label{1.8}
\end{equation}%
Next, theorem 3.3 of \cite{V1} and theorem 6 of Chapter 9 of \cite{V}
guarantee that $V(x,k,\nu )=u(x,k,\nu ),$ where the function $u(x,k,\nu )$
is the solution to the equation (\ref{1.4})-(\ref{1.4a}).

Using the representation (\ref{5.7}), we obtain 
\begin{equation}
u(x,k,\nu )=A(x,\nu )\exp (-ik\varphi (x,\nu ))+\dint\limits_{\varphi (x,\nu
)}^{\infty }\widehat{v}\left( x,t,\nu \right) \exp \left( -ikt\right)
dt,\quad x\in \Phi .  \label{1.9}
\end{equation}%
Denote 
\begin{equation*}
\partial _{t}^{j}\widehat{v}_{+}(x,\varphi (x,\nu ),\nu )=\lim_{t\rightarrow
\varphi ^{+}(x,\nu )}\partial _{t}^{j}\widehat{v}(x,t,\nu ),j=0,1.
\end{equation*}%
Integrating by parts in (\ref{1.9}), we obtain for $x\in \Phi $ 
\begin{equation}
u(x,k,\nu )=A(x,\nu )\exp (-ik\varphi (x,\nu ))+\frac{\exp (-ik\varphi
(x,\nu ))}{ik}\widehat{v}_{+}(x,\varphi (x,\nu ),\nu )  \notag  \label{1.10}
\end{equation}%
\begin{equation*}
+\frac{1}{ik}\dint\limits_{\varphi (x,\nu )}^{\infty }\widehat{v}_{t}\left(
x,t,\nu \right) \exp \left( -ikt\right) dt.\quad
\end{equation*}%
Hence,%
\begin{equation*}
u(x,k,\nu )=A(x,\nu )\exp (-ik\varphi (x,\nu ))+O\left( \frac{1}{k}\right)
,\quad k\rightarrow \infty ,\forall x\in \Phi ,\forall \nu \in S^{2}.
\end{equation*}%
This and (\ref{1.4a}) imply that for all $x\in \overline{\Omega },\nu \in
S^{2}$ 
\begin{equation}
u_{sc}(x,k,\nu )=A(x,y)\exp (ik\varphi (x,\nu ))-\exp (ikx\cdot \nu
)+O\left( \frac{1}{k}\right) ,\text{ }k\rightarrow \infty .  \label{6.1}
\end{equation}%
For every vector $\nu \in S^{2}$ denote $S^{+}(\nu )=\{x\in S:x\cdot \nu
>0\}.$

\section{Approximate reconstruction of the function $u_{sc}(x,k,\protect\nu %
) $ for all pairs $\protect\nu \in S^{2},x\in S^{+}(\protect\nu )$}

Ignoring the term $O\left( 1/k\right) $ in (\ref{6.1}) and using (\ref{100}%
), we obtain the following approximate formula 
\begin{equation}
f(x,k,\nu )=A^{2}(x,\nu )+1-2A(x,\nu )\cos \left[ k(\varphi (x,\nu )-x\cdot
\nu )\right] ,\forall \nu \in S^{2},\forall x\in S^{+}(\nu ).  \label{6.2}
\end{equation}

We now fix the pair $\nu \in S^{2},x\in S^{+}(\nu )$ and consider $f(x,k,y)$
as the function of $k$ for $k\geq k_{1}$, where $k_{1}\geq k_{0}$ is a
sufficiently large number. It is possible to figure out whether or not $%
\varphi (x,\nu )=x\cdot \nu .$ Indeed, it follows from (\ref{6.2}) that $%
\varphi (x,\nu )=x\cdot \nu $ if and only if $f(x,k,\nu )=const.$ for $k\geq
k_{1},$ i.e. if and only if $\partial _{k}f(x,k,\nu )=0,\forall k\geq k_{1}$.

Suppose now that $f(x,k,\nu )\neq const.$ for $k\geq k_{1}.$ It follows from
(\ref{6.2}) that there exists a number $k_{2}\geq k_{1}$ such that 
\begin{equation*}
f(x,k_{2},\nu )=\max_{k\geq k_{1}}f(x,k,\nu )=\left( A(x,\nu )+1\right) ^{2}.
\end{equation*}%
In particular, we find from here the number $A(x,\nu ),$ 
\begin{equation}
A(x,\nu )=\sqrt{f(x,k_{2},\nu )}-1.  \label{6.3}
\end{equation}%
By (\ref{6.2}) there exists a sequence $\left\{ k_{n}\left( x,\nu \right)
\right\} _{n=3}^{\infty }\subset \left( k_{2}\left( x,\nu \right) ,\infty
\right) $ such that 
\begin{equation*}
f(x,k_{n}\left( x,\nu \right) ,\nu )=\max_{k\geq k_{1}}f(x,k,\nu )\text{ and 
}k_{2}<k_{3}<...k_{n}<...
\end{equation*}%
Hence, 
\begin{equation*}
k_{3}\left( x,\nu \right) (\varphi (x,\nu )-x\cdot \nu )=k_{2}\left( x,\nu
\right) (\varphi (x,\nu )-x\cdot \nu )+2\pi .
\end{equation*}
Thus, we approximate the number $\varphi (x,\nu ),$%
\begin{equation}
\varphi (x,\nu )=x\cdot \nu +\frac{2\pi }{k_{3}\left( x,\nu \right)
-k_{2}\left( x,\nu \right) }.  \label{6.4}
\end{equation}

Thus, it follows from (\ref{6.1}) that formulae (\ref{6.3}) and (\ref{6.4})
provide us with an approximation of the function $u_{sc}(x,k,\nu )$ for
sufficiently large values of $k$ and for all pairs $\nu \in S^{2},x\in
S^{+}(\nu ).$ We have obtained this approximation only using the data (\ref%
{100}) for our inverse problem. We use the word \textquotedblleft
approximation" here because we got (\ref{6.2}) via ignoring the term $%
O\left( 1/k\right) $ in (\ref{6.1}). Next, we should reconstruct the
function $\beta \left( x\right) .$ This is done in section 7.

\section{ Linearization}

We assume in sections 6 and 7 that 
\begin{equation}
||\beta ||_{C(\overline{\Omega })}<<1.  \label{1.2}
\end{equation}%
Use the linearization method for $\varphi (x,\nu )$ proposed in \cite{LRV}
(see chapter 3 in this book and \cite{R1, R3}). Then, we represent solution
to problem (\ref{1.5a}) in the form 
\begin{equation}
\varphi (x,\nu )=\varphi _{0}(x,\nu )+\varphi _{1}(x,\nu ),  \label{2.5a}
\end{equation}%
where $\varphi _{0}(x,\nu )=x\cdot \nu $. By (\ref{1.0}) $n^{2}\left(
x\right) =1+2\beta \left( x\right) +\beta ^{2}\left( x\right) .$ By (\ref%
{1.2}) the term $\beta ^{2}\left( x\right) $ can be neglected in the latter
expression. Substituting the representation (\ref{2.5a}) in (\ref{1.5a}) and
neglecting terms $\left( \nabla \varphi _{1}\right) ^{2}$ and $\beta ^{2}$,
we find 
\begin{equation}
\nabla _{x}\varphi _{1}(x,\nu )\cdot \nabla _{x}\varphi _{0}(x,\nu )=\beta
(x),\quad \varphi _{1}(x,\nu )=0\>\text{ for all}\>\text{ }x\cdot \nu \leq
-B.  \label{2.5b}
\end{equation}%
Since $\nabla _{x}\varphi _{0}(x,\nu )=\nu $, then the left hand side of
equation (\ref{2.5b}) is the derivative in the direction $\nu $. Hence, 
\begin{equation}
\varphi _{1}(x,\nu )=\int\limits_{-x\cdot \nu -B}^{0}\beta (x+\nu \sigma
)d\sigma ,  \label{2.5c}
\end{equation}%
The right hand side of (\ref{2.5c}) is the integral over the segment $%
l\left( x,\nu \right) $ of a straight line and $\sigma $ is its arc length.
The segment $l\left( x,\nu \right) $ is stretched from the point $x$ in the
direction $-\nu $ and continues until reaching the point $\overline{x}\left(
x,\nu \right) =x-\left( x\cdot \nu +B\right) \nu \in \Sigma \left( \nu
\right) .$ Note that by (\ref{3.11}) $\overline{x}\left( x,\nu \right)
\notin \Omega .$

Let $\nu \in S^{2}$ and $x\in S^{+}(\nu )$. Denote by $y=y(x,\nu
)=x-2(x\cdot \nu )\nu $ the second intersection point of the straight line $%
\xi =x+\nu \sigma $ with $S$. Let $L\left( x,\nu \right) $ denotes the
segment of that straight line connecting points $x$ and $y\left( x,\nu
\right) .$ Since the function $\beta \left( x\right) =0$ outside of the ball 
$\Omega $ and since $S=\partial \Omega ,$ then equation (\ref{2.5c}) is
equivalent with%
\begin{equation}
\varphi _{1}(x,\nu )=\dint\limits_{L\left( x,\nu \right) }\beta \left( \xi
\right) d\sigma ,\forall \nu \in S^{2},\forall x\in S^{+}(\nu ).
\label{2.5d}
\end{equation}

\section{Reconstructions}

Formula (\ref{2.5d}) enables us to use two reconstruction procedures for
finding the function $\beta \left( x\right) .$ The first one is the inverse
2-d Radon transform, and the second one solves integral equations of the
Abel kind. As it was pointed out in Introduction, the idea of second
procedure goes back to the paper \cite{Cormack}. In both cases we can
reconstruct the function $\beta \left( x\right) $ separately in each 2-d
cross section $\left\{ x_{3}=a=const.\right\} $ of the ball $\Omega .$

For any number $a\in \mathbb{R}$ consider the plane $P_{a}=\left\{
x_{3}=a\right\} .$ Consider the disk $Q_{a}=\Omega \cap P_{a}$ and let the
circle $S_{a}=S\cap P_{a}$ be its boundary. Then the radius of this circle
is $B_{a}=\sqrt{B^{2}-a^{2}}.$ Clearly $Q_{a}\neq \varnothing $ for $a\in
\left( -B,B\right) $ and $Q_{a}=\varnothing $ for $\left\vert a\right\vert
\geq B.$ Denote $0_{a}=\left( 0,0,a\right) \in Q_{a}$ the orthogonal
projection of the origin on the plane $P_{a}.$ We have 
\begin{equation*}
\Omega =\dbigcup\limits_{a=-B}^{B}Q_{a},\partial \Omega
:=S=\dbigcup\limits_{a=-B}^{B}S_{a}.
\end{equation*}%
Denote 
\begin{eqnarray*}
S_{0}^{2} &=&\left\{ \nu =\left( \nu _{1},\nu _{2},\nu _{3}\right) \in
S^{2}:\nu _{3}=0\right\} , \\
S_{a}^{+}(\nu ) &=&\{x\in S_{a}:\,x\cdot \nu >0\},\forall \nu \in S_{0}^{2}.
\end{eqnarray*}

\subsection{Reconstruction using the inverse Radon transform}

\label{sec:6.1}

In this subsection we present the reconstruction formula based on the
inverse 2-d Radon transform. First, we parameterize $L\left( x,\nu \right) $
in the conventional way of the parametri\-zation of the Radon transform \cite%
{Nat}. For $\nu \in S_{0}^{2},x\in S_{a}^{+}(\nu ),$ let $m$ be the unit
normal vector to the line $L\left( x,\nu \right) $ lying in the plane $P_{a}$
and pointing outside of the point $0_{a}.$ Since we work now only with the
plane $P_{a},$ we discard the third component. Let $\alpha \in \left( 0,2\pi %
\right] $ be the angle between $m$ and the $x_{1}-$axis. Then $m=m\left(
\alpha \right) =\left( \cos \alpha ,\sin \alpha \right) $. Let $d$ be the
signed distance between $L\left( x,\nu \right) $ and the point $0_{a}$ (page
9 of \cite{Nat}). It is clear that there exists a one-to-one correspondence
between pairs $\left( x,\nu \right) $ and $\left( m\left( \alpha \right)
,d\right) ,d\left( x,\nu \right) \in \left( -B_{a},B_{a}\right) ,$%
\begin{equation}
\left( x,\nu \right) \Leftrightarrow \left( m\left( \alpha \right) ,d\right)
;\nu \in S_{0}^{2},x\in S_{a}^{+}(\nu );\alpha =\alpha \left( x,\nu \right)
\in \left( 0,2\pi \right] ,d=d\left( x,\nu \right) .  \label{1.80}
\end{equation}%
Hence, we can write 
\begin{equation}
L\left( x,\nu \right) =\left\{ z_{a}=\left( z_{1},z_{2},a\right) \in
Q_{a}:z\cdot m(\alpha )=d\right\} ,  \label{1.81}
\end{equation}%
where $z=\left( z_{1},z_{2}\right) \in \mathbb{R}^{2}$ and parameters $%
\alpha =\alpha \left( x,\nu \right) $ and $d=d\left( x,\nu \right) $ are
defined as in (\ref{1.80}).

Consider an arbitrary function $q=q\left( z\right) \in C^{2}\left(
P_{a}\right) $ such that $q\left( z\right) =0$ for $z\in $ $P_{a}\diagdown
Q_{a}.$ Hence, 
\begin{equation}
\dint\limits_{L\left( x,\nu \right) }q\left( z\right) d\sigma
=\dint\limits_{z\cdot m\left( \alpha \right) =d}q\left( z\right) d\sigma
,\>\forall x\in S_{a}(\nu ),\>\forall \nu \in S_{0}^{2},  \label{1.82}
\end{equation}%
where $\alpha =\alpha \left( x,\nu \right) ,s=s\left( x,\nu \right) $ are as
in (\ref{1.80}). In (\ref{1.82}) $\sigma $ is the arc length and the
parametrization of $L\left( x,\nu \right) $ is given in (\ref{1.81}).
Therefore, using (\ref{1.80})-(\ref{1.82}), we can define the 2-d Radon
transform $Rq$ of the function $q$ as 
\begin{equation}
\left( Rq\right) \left( x,\nu \right) =\left( Rq\right) \left( \alpha
,d\right) =\dint\limits_{z\cdot m\left( \alpha \right) =d}g\left( z\right)
d\sigma .  \label{1.83}
\end{equation}

Let $R^{-1}$ be the operator which is the inverse to the operator $R$ in (%
\ref{1.83}). The specific form of $R^{-1}$ is well known, see, e.g. \cite%
{Nat}. Hence, we do not present this form here for brevity. Then formulae (%
\ref{2.5d}) and (\ref{1.83}) imply that 
\begin{equation}
\beta \left( z,a\right) =R^{-1}\left( \varphi _{1}(x,\nu )\right) \left(
z,a\right) ,\forall a\in \left( -B,B\right) ,\forall z\in Q_{a}.
\label{1.84}
\end{equation}%
By (\ref{6.4}) and (\ref{2.5a}) the function $\varphi _{1}(x,\nu )$ is
approximately known for all $\nu \in S^{2}$ and for all $x\in S_{a}^{+}(\nu
).$ Thus, formula (\ref{1.84}) completes our reconstruction process via the
inverse Radon transform.

\subsection{Reconstruction via solutions of integral equations of the Abel
type}

Consider now again an arbitrary number $a\in \left( -B,B\right) $ and an
arbitrary pair $\nu \in S_{0}^{2},x\in S_{a}^{+}(\nu ).$ Then the point $%
y\left( x,\nu \right) =\left( y_{1},y_{2},a\right) \in S_{a}.$ Given the
above pair $x,\nu ,$ there exists unique point $y\left( x,\nu \right) \in
S_{a}.$ And vice versa: given a point $y\in S_{a}$ and a vector $\nu \in
S_{0}^{2},$ there exists unique point $x\in S_{a}^{+}(\nu )$ such that the
segment $L\left( x,\nu \right) $ passes through points $x$ and $y$. Hence,
we denote below $M\left( x,y\right) $ the segment of the straight line
passing through points $x,y\in S_{a}.$ Clearly the set $\left\{ M\left(
x,y\right) \right\} _{x,y\in S_{a}}$ coincides with the set $\left\{ L\left(
x,\nu \right) \right\} _{\nu \in S_{0}^{2},x\in S_{a}^{+}(\nu )},$ i.e. $%
\left\{ M\left( x,y\right) \right\} _{x,y\in S_{a}}=\left\{ L\left( x,\nu
\right) \right\} _{\nu \in S_{0}^{2},x\in S_{a}^{+}(\nu )}.$ Keeping this in
mind, we rewrite (\ref{2.5d}) as 
\begin{equation}
\psi \left( x,y\right) =\dint\limits_{M\left( x,y\right) }\beta \left( \xi
\right) d\sigma ,\forall x,y\in S_{a},\forall a\in \left( -B,B\right) ,
\label{1.900}
\end{equation}%
where the function $\psi (x,y)$ is constructed from the function $\varphi
_{1}(x,\nu )$ in an obvious manner, i.e. for each pair $x,y\in S_{a}$ we
find the vector $\nu \in S_{0}^{2}$ such that $x\in S_{a}^{+}(\nu )$ and $%
y=y\left( x,\nu \right) ,$ and then set $\psi (x,y)=$ $\psi (x,y\left( x,\nu
\right) )=\varphi _{1}(x,\nu ).$

In the plane $P_{a}$ we introduce polar coordinates $r,\varphi $ of the
variable point $\xi =(\xi _{1},\xi _{2})$ as $\xi _{1}=r\cos \phi ,\xi
_{2}=r\sin \phi $. We characterize $M(x,y)$ by the polar coordinates $(\rho
,\alpha )$ of its middle point $(x+y)/2$. Hence, $\left\vert x-y\right\vert
=2\sqrt{B^{2}-a^{2}-\rho ^{2}}.$ Change variables in the integral (\ref%
{1.900}) as 
\begin{equation*}
\sigma \Longleftrightarrow r,\sigma =\sqrt{B^{2}-a^{2}-\rho ^{2}}-\sqrt{%
r^{2}-\rho ^{2}},
\end{equation*}%
Then $d\sigma =-rdr/\sqrt{r^{2}-\rho ^{2}}$. The equation of $M(x,y)$ can be
rewritten as 
\begin{equation}
\phi =\alpha +(-1)^{j}\arccos \frac{\rho }{r},\quad j=1,2,\quad r\geq \rho ,
\label{1.90}
\end{equation}%
where $j=1$ corresponds to the part of the segment $M(x,y)$ between points $%
(x+y)/2$ and $x$, and $j=2$ to the rest of $M(x,y)$.

Obviously, there exists a one-to-one correspondence, up to the symmetry
mapping $(x,y)\Leftrightarrow (y,x)$ between pairs $x,y\in S_{a}$ and $(\rho
,\alpha )\in \left( 0,R\right) \times \left( 0,2\pi \right) $. Denote

\begin{equation*}
\beta (r\cos \phi ,r\sin \phi ,a)=\widetilde{\beta }(r,\varphi ,a)\text{ and 
}\widetilde{\psi }(\rho ,\alpha ,a)=\psi (x,y).
\end{equation*}%
Using (\ref{1.90}), we rewrite equation (\ref{1.900}) as 
\begin{equation}
\sum\limits_{j=1}^{2}\int\limits_{\rho }^{B_{a}}\widetilde{\beta }(r,\alpha
+(-1)^{j}\arccos \frac{\rho }{r},a)\frac{r}{\sqrt{r^{2}-\rho ^{2}}}dr=%
\widetilde{\psi }(\rho ,\alpha ,z).  \label{1.20}
\end{equation}%
Represent functions $\widetilde{\beta }(r,\varphi ,a)$ and $\widetilde{\psi }%
(\rho ,\alpha ,a)$ via Fourier series, 
\begin{equation*}
\widetilde{\beta }(r,\varphi ,a)=\sum\limits_{n=-\infty }^{\infty }%
\widetilde{\beta }_{n}(r,a)\exp (in\phi ),
\end{equation*}%
\begin{equation*}
\widetilde{\psi }(\rho ,\alpha ,z)=\sum\limits_{n=-\infty }^{\infty }%
\widetilde{\psi }_{n}(\rho ,a)\exp (in\alpha ).
\end{equation*}%
Multiplying both sides of (\ref{1.20}) by $\exp (-in\alpha )/(2\pi )$ and
integrating with respect to $\alpha \in \left( 0,2\pi \right) $, we obtain
for all $n=0,\pm 1,\pm 2,\ldots $ 
\begin{equation}
\int\limits_{\rho }^{B_{a}}\widetilde{\beta }_{n}(r,a)\cos \Big(n\arccos 
\frac{\rho }{r}\Big)\>\frac{rdr}{\sqrt{r^{2}-\rho ^{2}}}=\widetilde{\psi }%
_{n}(\rho ,a),\rho \in \left( 0,B_{a}\right)  \label{1.23}
\end{equation}%
This is the integral equation of the Abel type. To solve equation (\ref{1.23}%
), we apply first the operator $A$ to both sides of (\ref{1.23}), where 
\begin{equation*}
A\left( h\left( \rho \right) \right) \left( \omega \right) =\frac{1}{\pi }%
\int\limits_{s}^{B_{a}}\frac{h(\rho )\rho \,d\rho }{\sqrt{\rho ^{2}-\omega
^{2}}},\text{ }\omega \in \left( 0,B_{a}\right) .
\end{equation*}%
Changing the limits of the integration, we obtain%
\begin{equation}
\frac{1}{\pi }\dint\limits_{\omega }^{B_{a}}\widetilde{\beta }_{n}(r,a)\left[
\dint\limits_{\omega }^{r}\frac{\rho }{\sqrt{\rho ^{2}-\omega ^{2}}\cdot 
\sqrt{r^{2}-\rho ^{2}}}\cos \Big(n\arccos \frac{\rho }{r}\Big)d\rho \right]
dr  \label{102}
\end{equation}%
\begin{equation*}
=A\left( \widetilde{\psi }_{n}(\rho ,a)\right) \left( \omega \right) .
\end{equation*}%
Change variables in the inner integral (\ref{102}) as%
\begin{equation*}
\rho \Leftrightarrow \theta ,\rho ^{2}=\omega ^{2}\cos ^{2}\left( \theta
/2\right) +r^{2}\sin ^{2}\left( \theta /2\right) .
\end{equation*}%
Then%
\begin{equation*}
2\rho d\rho =\left( r^{2}-\omega ^{2}\right) \sin \theta \cos \theta d\theta
,
\end{equation*}%
\begin{equation*}
\sqrt{\rho ^{2}-\omega ^{2}}\cdot \sqrt{r^{2}-\rho ^{2}}=\left( r^{2}-\omega
^{2}\right) \sin \theta \cos \theta .
\end{equation*}%
Hence, equation (\ref{102}) can be rewritten as%
\begin{equation}
\int\limits_{\omega }^{B_{a}}\widetilde{\beta }_{n}(r,a)Q_{n}(r,\omega
)dr=2A\left( \widetilde{\psi }_{n}(\rho ,a)\right) \left( \omega \right) ,
\label{1.26}
\end{equation}%
\begin{equation*}
Q_{n}(r,\omega )=\frac{1}{\pi }\int\limits_{0}^{\pi }\cos \left( n\arccos 
\frac{\sqrt{r^{2}\cos ^{2}(\theta /2)+\omega ^{2}\sin ^{2}(\theta /2)}}{r}%
\right) d\theta .
\end{equation*}%
We have $Q_{n}(\omega ,\omega )=1$. Hence, differentiating (\ref{1.26}) with
respect to $\omega $, we obtain Volterra integral equation of the second
kind 
\begin{equation}
\widetilde{\beta }_{n}(\omega ,a)-\int\limits_{\omega }^{B_{a}}\widetilde{%
\beta }_{n}(r,a)T_{n}(r,\omega )dr=-\frac{\partial }{\partial \omega }\left[
2A\left( \widetilde{\psi }_{n}(\rho ,z)\right) \left( \omega \right) \right]
,\text{ }\omega \in \left( 0,B_{a}\right) ,  \label{1.28}
\end{equation}%
\begin{equation}
T_{n}(r,\omega )=\frac{n}{\pi \sqrt{r^{2}-\omega ^{2}}}\times  \label{1.29}
\end{equation}%
\begin{equation*}
\int\limits_{0}^{\pi }\left[ \sin \left( n\arccos \left( \frac{\sqrt{%
r^{2}\cos ^{2}(\theta /2)+\omega ^{2}\sin ^{2}(\theta /2)}}{r}\right)
\right) \frac{\sin (\theta /2)}{\sqrt{r^{2}\cos ^{2}(\theta /2)+\omega
^{2}\sin ^{2}(\theta /2)}}\right] d\theta .
\end{equation*}%
It follows from (\ref{1.29}) that the kernel of integral equation (\ref{1.28}%
) has the form%
\begin{equation*}
T_{n}(r,\omega )=\frac{\widetilde{T}_{n}(r,\omega )}{\sqrt{r^{2}-\omega ^{2}}%
},
\end{equation*}%
where the function $\widetilde{T}_{n}(r,\omega )$ is continuous for $0\leq
\omega \leq r\leq B_{a}$. Therefore, it follows from the theory of Volterra
integral equations of the second kind that for each $a\in \left( -B,B\right) 
$ there exists a solution $\widetilde{\beta }_{n}(\omega ,a)\in C\left[
0,B_{a}\right] $ of equation (\ref{1.28}) and this solution is unique.
Furthermore, it is well known from that theory that equation (\ref{1.28})
can be solved iteratively as%
\begin{equation}
\widetilde{\beta }_{n}^{0}(\omega ,a)=-\frac{\partial }{\partial \omega }%
\left[ M\left( \widetilde{\psi }_{n}(\rho ,a)\right) \left( \omega \right) %
\right] ,  \label{1.30}
\end{equation}%
\begin{equation}
\widetilde{\beta }_{n}^{k}(\omega ,a)=\int\limits_{\omega }^{\rho _{0}}%
\widetilde{\beta }_{n}^{k-1}(r,a)T_{n}(r,\omega )dr-\frac{\partial }{%
\partial \omega }\left[ M\left( \widetilde{\psi }_{n}(\rho ,a)\right) \left(
\omega \right) \right] ,k=1,2,...  \label{1.32}
\end{equation}%
and this process converges in the space $C\left[ 0,\rho _{0}\right] $ to the
solution $\widetilde{\beta }_{n}(\omega ,a)$ of equation (\ref{1.28}).
Formulae (\ref{1.30}) and (\ref{1.32}) finish our second reconstruction
procedure.

\begin{center}
\textbf{Acknowledgments}
\end{center}

The work of the first author was supported by the US Army Research
Laboratory and US Army Research Office grant W911NF-15-1-0233 as well as by
the Office of Naval Research grant N00014-15-1-2330. The work of the second
author was partially supported by the Russian Foundation for Basic Research,
grant No. 14-01-00208.\hfill

\end{document}